\documentclass[conference]{IEEEtran}
\usepackage{cite}
\usepackage{amsmath,amssymb,amsfonts}
\usepackage{algorithmic}
\usepackage{graphicx}
\usepackage{textcomp}
\usepackage{xcolor}
\usepackage[nolist]{acronym}
\usepackage{comment}
\usepackage{mathrsfs,bm} 
\usepackage{scalerel}
\usepackage[caption=false]{subfig}
\usepackage{tikz}
\usepackage{pgfplots}
\usepackage{tabu}
\usepackage{flushend}

\usepackage{mathtools}
\mathtoolsset{showonlyrefs}

\DeclareMathOperator{\sinc}{sinc}
\DeclareMathOperator{\diag}{diag}

\newcommand{\ee}{\mathrm{e}}

\definecolor{pant24PeachFuzz}{RGB}{255,190,152}

\usepackage{xcolor}
\usepackage{soul}
\usepackage{svg}

\IEEEoverridecommandlockouts
\begin{document}

\title{A low-PAPR Pilot Design and Optimization for OTFS Modulation}

\author{Davide~Bergamasco, Federico~Clazzer, Andrea~Munari, and Paolo~Casari
\thanks{D.~Bergamasco, F.~Clazzer and A.~Munari are with the Inst.\ of Communications and Navigation, German Aerospace Center (DLR), Wessling, Germany (e-mail: \{davide.bergamasco, federico.clazzer,  andrea.munari\}@dlr.de).}%
\thanks{P.~Casari is with DISI, University of Trento,  Italy (e-mail: paolo.casari@unitn.it). D.~Bergamasco is also with DISI.}%
\thanks{D. Bergamasco, F. Clazzer and A. Munari acknowledge the financial support by the Federal Ministry of Education and Research of Germany in the programme of ``Souver{\"a}n. Digital. Vernetzt.'' Joint project 6G-RIC, project identification number: 16KISK022. The work of P.~Casari was partally supported by the European Union -- Next Generation EU -- PNRR, Mission 4 Component 2, Investment 1.3 -- PE RESTART Spoke~6 - Project EMBRACE (PE00000001, CUP E63C22002070006).}%
}

\maketitle
\thispagestyle{empty}

\begin{abstract}
\Ac{OTFS} modulation has been proposed recently as a new waveform in the context of doubly-selective multi-path channels. This article proposes a novel pilot design that improves \ac{OTFS}'s \ac{SE} while reducing its \ac{PAPR}. Instead of adopting an embedded data-orthogonal pilot for channel estimation, our scheme relies on Chu sequences superimposed to data symbols. We optimize the construction by investigating the best energy split between pilot and data symbols. Two equalizers, and an iterative channel estimation and equalization procedure are considered. We present extensive numerical results of relevant performance metrics, including the normalized mean squared error of the estimator, bit error rate, \ac{PAPR} and \ac{SE}. Our results show that, while the embedded pilot scheme estimates the channel more accurately, our approach yields a better tradeoff by achieving much higher spectral efficiency and lower \ac{PAPR}.
\end{abstract}

\section{Introduction}
\acresetall
In the context of broadband communication systems, high-mobility wireless channels introduce severe distortions to transmitted signals by spreading their power across the time and the frequency domains. In these scenarios, \ac{OTFS} modulation~\cite{OTFS_original} has recently attracted a lot of attention thanks to its capability to exploit the appealing properties offered by the \ac{DD} representation of the channel. 
\ac{OTFS} can be seen as a digital multi-carrier system that orthogonally multiplexes $M N$ information symbols, with $M, N \in \mathbb{N} \setminus\{0\}$, in the \ac{DD} domain. Any non-trivial multi-path channel destroys this orthogonality at the receiver, introducing the need for equalization to cancel the \ac{ICI} and restore the transmitted data. One unique property of \ac{OTFS} is that, when properly designed, the interaction between any of the $M N$ sub-carriers and the channel is the same. More formally, when the \emph{crystallization condition}~\cite{cristallyzation} is fulfilled, successive replicas of the channel response in the \ac{DD} domain are guaranteed to be free of aliasing. Hence, under this condition, it is sufficient to estimate the channel behavior once for all sub-carriers. 

Along this line, many channel estimation algorithms have been proposed in recent years, e.g.,~\cite{EP, health, Zhiqiang2022, Himanshu2022, Weijie2021, S1D}. In~\cite{EP}, a design considering one single pilot allocated within an \ac{OTFS} frame is considered. A sufficient guard band of empty symbols surrounds the pilot so that, even after propagating through the channel, the received pilot signal remains free of interference  from data. In~\cite{health}, the design is extended to \ac{MIMO} channels, and in~\cite{Zhiqiang2022} a sparse Bayesian recovery is introduced to estimate the channel paths.
These approaches preserve data-pilot orthogonality at the receiver, favoring accurate channel estimation and thus enabling effective data equalization. 
However, reserving resources for pilot and guard symbols in the \ac{OTFS} frame reduces the \ac{SE} of the scheme.
To improve this metric, a different design strategy that embraces data-pilot interference has also been explored. 
This idea has been firstly presented in the context of \ac{OTFS} in~\cite{Himanshu2022}, where a bi-dimensional pilot sequence is superimposed to data symbols so that both contributions cover one entire \ac{OTFS} frame. A related construction has been proposed in~\cite{Weijie2021}, where a single pilot is placed within an \ac{OTFS} frame full of data symbols. In~\cite{S1D}, instead, the pilot sequence only covers one of the columns along the delay dimension of the \ac{OTFS} frame, while the data is still allowed to span the entire frame.

Motivated by use cases with tight link budgets and limited energy resources, such as in satellite-enabled \ac{IoT} applications, we aim to devise an \ac{OTFS} scheme with low \ac{PAPR} and high \ac{SE}. Thus, we focus on a construction that overlaps pilots and data. Observing that a pilot sequence covering one entire column of the \ac{OTFS} frame minimizes the pilot contribution to the \ac{PAPR}, we proceed along the same line as~\cite{S1D}. However, unlike~\cite{S1D}, we optimize the energy split among data and pilots and do not consider the number of multi-path channel components to be known. Additionally,  we propose the use of Chu sequences~\cite{CHU}, which are specifically suited to channel estimation via sparse recovery algorithms. 
Moreover, we present an iterative channel estimation and equalization procedure that considerably boosts the performance of non-orthogonal designs. Finally, we numerically evaluate the performance of the proposed approach in terms of the \ac{NMSE} of the estimator, \ac{BER}, \ac{PAPR}, and \ac{SE}. We compute these metrics also for an orthogonal data-pilot design and discuss the involved trade-offs together with the role of orthogonality.

\section{System model}

\begin{figure*}
    \centering
    \includegraphics[width=0.9\textwidth]{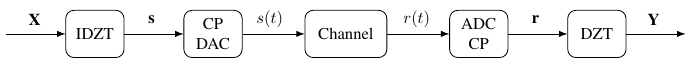}
    \caption{IDZT-OTFS communication pipeline.}
    \label{fig:pipeline}
\end{figure*}

In this section, we describe the \ac{OTFS} communication model, as shown in Fig.~\ref{fig:pipeline}.

\vspace{-6pt}
\subsection{Transmitter}
The input signal $\textbf{X} \in \mathbb{C}^{M\times N}$ consists of $M N$ complex symbols that represent the available degrees of freedom at the input of the communication chain. These are used to transmit data and pilot within each frame, where the two contributions are respectively contained in the matrices $\textbf{X}_d$ and $\textbf{X}_p$ so that $\textbf{X} = \textbf{X}_d + \textbf{X}_p$.  
The signal $\textbf{X}$ is represented in the \ac{DD} domain and,  by means of the \ac{IDZT}, it can be converted into its discrete-time representation ${\bm{s} \in \mathbb{C}^{NM\times 1}}$ as
\begin{equation}
 \label{eq:IDZT}
          \bm{s}[t] = \bm{s}[\ell + nM] = \frac{1}{\sqrt N} \sum^{N-1}_{k=0} \textbf{X}[\ell,k]\ee{}^{j\frac{2 \pi}{N}nk} 
\end{equation}
with $\ell \in \{0, .. \,,M-1\}$, $n \in \{0,.. \, , N-1\}$ and $t=\ell+nM$. The transformation shown in \eqref{eq:IDZT} is performed through the IDZT-block depicted in Fig.~\ref{fig:pipeline} and can be equivalently computed as
\begin{equation}
\label{eq:IDZT_mat}
        \bm{s} = \text{IDZT}\{\textbf{X}\}=\text{vec}(\textbf{X}\textbf{F}_N^H) = (\textbf{F}_N^H \otimes \textbf{I}_{M})\bm{x} 
\end{equation}
where $\bm{x} = \text{vec}(\textbf{X})$, is obtained by stacking all columns of $\textbf{X}$, $\textbf{F}_N$ is the \ac{DFT} matrix of size $N$ and $\otimes$ represents the Kronecker product. To avoid inter-frame interference, we make use of a \ac{CP} at the frame level, leading to the so-called \ac{RCP} version of \ac{OTFS}~\cite{GeneralI/O}. This is implemented by copying and pre-pending the last $L$ samples of $\bm{s}$ so that the overall signal is composed of $MN+L$ symbols. The frame duration is given as $T_{f}=(MN+L)T_s$, where $T_s$ is the symbol time. Finally, the resulting digital signal is converted to the baseband continuous signal $s(t)$, up-converted to radio frequency, and transmitted over the channel. We assume the use of rectangular pulse-shaping waveforms. 

\vspace{-6pt}
\subsection{Channel}
We consider a $P$-sparse multi-path channel with \ac{AWGN}. Each propagation component has its own complex gain, propagation delay and Doppler shift represented by $h_i$, $\tau_i$ and $\nu_i$ respectively, for $i \in \{1, .. \, ,P \}$.\footnote{These parameters are assumed to be constant for one frame duration as commonly done in \ac{OTFS} literature. This can be explained since the \ac{DD} domain requires substantial change of moving speed or propagation delays to change its representation.}
The input-output relation in the time domain is given as
\begin{equation}
\label{eq:iotime}
r(t) = \sum^{P}_{i=1} h_i \ee{}^{j2\pi\nu_i(t-\tau_i)}s(t-\tau_i) + n(t) 
\end{equation}
where $n(t) \sim  \mathcal{CN}(0,\sigma^2)$ is AWGN noise. The quantities $\nu_i$ and $\tau_i$ are expressed as a function of the delay and Doppler resolutions of the system
\begin{equation}
\label{eq:resolution}
\tau_i = \ell_i\cdot T_s \quad \text{and} \quad \nu_i = \frac{k_i}{T_f}= \frac{k_i}{NMT_s}\;.
\end{equation}
The delay spread and maximum Doppler shift are represented as $\ell_{M}=\max_{1\leq i\leq P}\left(\ell_i\right)$ and $k_{M}=\max_{1\leq i\leq P}\left(|k_i|\right)$. Using the sampling theorem it is possible to represent the band-limited signal $s(t)$ in terms of its samples~\cite{TSE} and by combining \eqref{eq:iotime} and \eqref{eq:resolution} we get, after minor manipulations
\begin{equation}
\label{eq:r[t]}
r[t] = \sum_{\ell=0}^{M-1} s[t-\ell] \sum^{P}_{i=1}h_i \ee{}^{j\frac{2\pi}{NM}k_{i} (t - \ell_{i}) }  \cdot \sinc\left( t - \ell_i\right)+n[t]
\end{equation}
where $\sinc(x) =\sin(\pi x)/(\pi x)$ and $\sinc(0)=1$. By assuming that ${\ell_i\in \mathbb{N}}$ and ${k_i \in \mathbb{Z}}$, meaning that $\tau_i$ and $\nu_i$ are integer multiples of the delay and Doppler resolutions, respectively,~\eqref{eq:r[t]} can be simplified as
\begin{equation}
\label{eq:r[t]2}
r[t] = \sum^{P}_{i=1}h_i \ee{}^{j\frac{2\pi}{NM}k_{i} (t - \ell_{i}) }  \cdot s[t - \ell_i]+n[t].
\end{equation}
We can now represent the input-output relation in matrix form as \cite[Section. 3]{GeneralI/O}
\begin{equation}
\label{eq:timeIO}
        \bm{r} = \textbf{G}\bm{s} + \bm{n} 
\end{equation}
where the time-domain channel matrix $\textbf{G} \in \mathbb{C}^{NM \times NM}$ is
\begin{equation}
\label{eq:G}
        \textbf{G} = \sum^{P}_{i=1}h_i \ee{}^{-j\frac{2\pi}{NM}k_{i}\ell_{i}}\boldsymbol{\Delta}^{k_i}\boldsymbol{\Pi}^{\ell_i} 
\end{equation}
with $\boldsymbol{\Delta}^{k_i} = \diag{}[\ee{}^{\frac{j2\pi}{NM}k_{i}\cdot(0)}, ... \,, \ee{}^{\frac{j2\pi}{NM}k_{i}\cdot(NM-1)} ]$, and where $\boldsymbol{\Pi}$ is the cyclic shift matrix of size $NM \times NM$
\begin{equation}
        \boldsymbol{\Pi} = \begin{bmatrix}
    0 & \cdots & 0 & 1\\
    1 & \cdots & 0 & 0\\
    \vdots & \ddots & \vdots & \vdots \\
    0 & \cdots & 1 & 0
\end{bmatrix}.
\end{equation}

\vspace{-9pt}
\subsection{Receiver}
The continuous-time signal $r(t)$ at the output of the channel is converted to baseband and sampled at frequency $f_c = \frac{1}{T_s}$ using an \ac{ADC}. After \ac{CP} removal and through the \ac{DZT}-block, the receiver represents the signal $\bm{r}$ in the \ac{DD} domain as
\begin{equation}
\label{eq:DZT_mat}
        \textbf{Y} = \text{DZT}\{\bm{r}\}=\textbf{F}_N \text{vec}^{-1}_{MN}(\bm{r}) 
\end{equation}
where $\text{vec}^{-1}_{ij}(\bm{x})$ reshapes a vector of length $ij$ into a matrix of size $i \times j$. By vectorizing the matrix $\textbf{Y}$ column-wise we define the vector 
\begin{equation}
\label{eq:yvec}
\bm{y} = \text{vec}(\textbf{Y}) = (\textbf{F}_N \otimes \textbf{I}_{M})\bm{r}.
\end{equation}
Considering \eqref{eq:timeIO} we multiply both sides of the equality by $\textbf{F}_N \otimes \textbf{I}_{M}$. Then, by combining the obtained result with \eqref{eq:IDZT_mat} and \eqref{eq:yvec}, we obtain the vectorized \ac{DD} input-output relation
\begin{equation}
\label{eq:ddIO}
\bm{y} = \textbf{H}\bm{x} + \bm{w} 
\end{equation}
where the \ac{DD} channel matrix $\textbf{H} \in \mathbb{C}^{NM \times NM}$ is given as
\begin{equation}
\label{eq:H}
\textbf{H}=(\textbf{F}_N \otimes \textbf{I}_{M})\textbf{G} (\textbf{F}_N^H \otimes \textbf{I}_{M}) 
\end{equation}
and $\bm{w}=(\textbf{F}_N \otimes \textbf{I}_{M})\bm{n} \in \mathbb{C}^{NM\times 1}$ represents the noise contribution in the transformed domain.

\section{Pilot design and estimation procedure}
\label{sec:pilots}
To minimize the overhead of the pilot in terms of \ac{DD} resources, a superimposed data-pilot configuration is considered. One important metric that drives our pilot signal construction is the \ac{PAPR}
\begin{equation}
    \text{PAPR} \left( f[t] \right) = \frac{\max\left( |f[t]|^2\right)}{ \mathbb{E} \left\{ f[t] \right\} } \; .
\end{equation}
This quantity provides a measure of how much the envelope of a signal $f[t]$ has uneven power peaks. 
Ideally, we would like to find the arrangement of pilot symbols in \ac{DD} domain that leads to a constant-power time signal.
The \ac{IDZT} modulator performs row-wise \ac{IFFT} of length $N$, thus recombining the energy contained in each row of $\textbf{X}$. A possible pilot construction to minimize the \ac{PAPR} is to equally spread the pilot energy among all the rows of $\textbf{X}$~\cite{Farhang23}. This can be done by considering a pilot sequence composed of $M$ equal-power symbols placed on a single column of $\textbf{X}$, thus spanning the entire delay dimension.

A construction of this type was considered in~\cite{S1D} where the authors make use of \ac{ZC} sequences~\cite{ZC}. These sequences are well suited for channel estimation from a \ac{PAPR} point of view given that all elements have the same amplitude. Moreover, they also have perfect periodic autocorrelation, which is equal to zero for any non-zero cyclic shift. This last property will be leveraged during the channel estimation phase to avoid pilot self-interference. When the former two properties are simultaneously met, the sequence is said to be \ac{CAZAC}. 

One downside of \ac{ZC} sequences is that the perfect autocorrelation property is only fulfilled for prime-valued lengths of the sequence. 
This places an additional requirement on the parameter $M$ that represents the number of rows of the matrix $\textbf{X}$, and affects several other aspects of the communication system.  For example, in order to fulfill the \emph{crystallization condition} we have to ensure that $\ell_M < M$ and $2k_M + 1 \le  N$. The dimensions of the DD matrix cannot be independently selected since their product, namely $NM$,  represents the number of orthogonal sub-carriers inside each OTFS frame, which is equal to  $(T_f-LT_s)B$ where $B=\frac{1}{T_s}$. Overall, by fixing the time-frequency resources allocated for the transmission of a frame, and thus defining its duration and bandwidth, the quantity $MN$ becomes fixed. At this point, any choice on the selection of $M$ implies the selection of $N$ and vice versa.  Ideally, we would like to have $N= 2^i$ with $i \in \mathbb{N}$ so that the $M$ $N$-\ac{IFFT}s operations inside the \ac{IDZT} and \ac{DZT} blocks can be more efficiently implemented. In general, the values of $N$ and $M$ might also play other roles once a specific pilot design and equalization method are selected.
 

\vspace{-6pt}
\subsection{Proposed pilot design}
\label{subsec:pilotsdesign}
\begin{figure}
    \centering
    \includesvg[width=\linewidth]{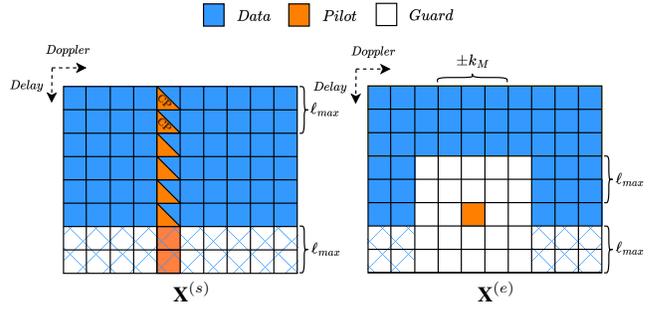}
    \caption{Our proposed \ac{S1D} data-pilot design (left) vs the embedded pilot construction from~\cite{EP} (right). The shaded elements represent data frame locations that would not affect the channel estimation procedure but are kept empty to enable the use of simpler equalizers.}
     \label{fig:pilots}
\end{figure}
In order to relax additional constraints on the parameter $M$, we propose alternatively the use of Chu sequences~\cite{CHU}, which always fulfill the \ac{CAZAC} property for any length and are defined as
\begin{align}
\bm{a}_p[y] = \left\{
\begin{aligned}
& \ee{}^{j\frac{\pi W}{K}y^2} && \text{if } K \text{ is even} \\
& \ee{}^{j\frac{\pi W}{K}y(y+1)} && \text{otherwise}
\end{aligned}
\right.
\end{align}
where $K$ is the sequence length, $W$ is an integer prime to $K$ and $y \in \{0,...\,,K-1 \}$. The proposed pilot configuration, along with the data, is shown on the left side of Fig.~\ref{fig:pilots}, where the Chu sequence is placed along one of the columns of the matrix $\textbf{X}_p^{(s)}$ and it is superimposed with data symbols. For the proposed \acf{S1D} approach we have that
\begin{equation}
     \textbf{X}^{(s)} =  \textbf{X}^{(s)}_p+  \textbf{X}^{(s)}_d
\end{equation}
where the matrix $\textbf{X}^{(s)}_d \in \mathbb{C}^{M\times N}$ contains $(M-\ell_{M})N$ data symbols in the first $M-\ell_{M}$ rows. The length of the employed Chu sequence is equal to $M-\ell_{M}$ to enable the insertion of a pilot \ac{CP} of length $\ell_{M}$ in front of it. This is done to preserve the autocorrelation property of the sequence even after the channel effect and will be formally justified in the last part of this paragraph. Hence, the pilot vector embedded into the matrix $\textbf{X}_p^{(s)}$ is given as
\begin{equation}
\begin{split}
\label{eq:CPpilot}
    \bm{a}_p^{CP} = &\,[\;\underbrace{a_p[M-2\ell_{M}], \ldots, a_p[M-\ell_{M}-1]}_{\begin{matrix}
        \bm{a}^{CP}
    \end{matrix}},\\
    &\,\underbrace{a_p[0], \ldots,a_p[M-\ell_{M}-1]}_{\begin{matrix}
        \bm{a}_p
    \end{matrix}}\;] 
\end{split}
\end{equation}
where $\bm{a}^{CP}\in \mathbb{C}^{\ell_{M}}$ is the pilot-\ac{CP}. We now clarify why adding the \ac{CP} to the pilot is important. Consider the discrete \ac{DD} input-output relation that can be derived by applying the \ac{DZT} to \eqref{eq:r[t]2} \cite[Chapter 5]{OTFSbook}
\begin{align}
\textbf{Y}[\ell,k] & = \sum\limits^{P}_{i=1}h_i \ee{}^{\frac{j 2\pi }{MN}k_{i} (\ell - \ell_{i})} \nonumber \\
 &\cdot \textbf{X}[ (\ell - \ell_{i})_M,(k-k_i)_N]\ee{}^{\frac{j 2\pi }{N}(k - k_{i}) \left\lfloor \frac{\ell - \ell_{i}}{M} \right\rfloor} \; . \label{eq:IODD}
\end{align}

From \eqref{eq:IODD}, we observe that the DD domain is quasi-periodic, i.e., it is periodic up to phase rotations. Each path of the channel scales and shifts the transmitted signal $\textbf{X}$ along both dimensions according to the complex gains $h_i$ and the \ac{DD} shifts $\ell_i$ and $k_i$. Within every frame, the channel response is completely characterized by the parameters $h_i,\ell_i,k_i$, and~$P$. By neglecting for the moment the presence of data symbols and noise, the receiver will only observe $P$ shifted and scaled replicas of the transmitted pilot. When the perfect autocorrelation property of the sequence is fulfilled, one can uniquely identify all the channel parameters by simply checking how many different pilot replicas can be found within the matrix $\textbf{Y}$. Recall that, when the \ac{CAZAC} property is satisfied, the autocorrelation is ideal. 
Unfortunately, the phase terms in \eqref{eq:IODD} have to be compensated, otherwise the \ac{CAZAC} property of the pilot is lost. Nevertheless, it can be shown that the only problematic term is $\exp \left({\frac{j 2\pi }{N}(k - k_{i}) \left\lfloor \frac{\ell - \ell_{i}}{M} \right\rfloor} \right)$, which only affects the first $\ell_{M}$ rows of $\textbf{Y}$ since $\ell_i\leq \ell_{M}$. Therefore, for channel estimation purposes, we solve this problem at the receiver by considering only the last $(M-\ell_{M})$ rows of $\bm{Y}$. 
Note that, the channel-induced shifts on the transmitted pilot sequence can be seen as circular shifts only when the period of the \ac{DD} domain along the delay dimension has the same size as the transmitted sequence. However, since at the receiver we are only processing the last $(M-\ell_{M})$ rows of the DD matrix, we have to properly select the length of the pilots. 
Hence, to enforce the periodic behavior of the channel-induced shifts on a period of length $(M-\ell_{M})$ we have to introduce the pilot-CP.

We are interested in comparing our pilot construction with what is commonly considered in \ac{OTFS} literature. Hence, we consider the embedded pilot design of~\cite{EP} (see Fig.~\ref{fig:pilots}) 
\begin{equation}
     \textbf{X}^{(e)} =  \textbf{X}^{(e)}_p+  \textbf{X}^{(e)}_d
\end{equation}
where the matrix $\textbf{X}^{(e)}_p \in \mathbb{C}^{M\times N}$ contains one unique pilot symbol, and $\textbf{X}^{(e)}_d \in \mathbb{C}^{M\times N}$ contains data symbols in all positions except the last $\ell_{M}$ rows and a rectangle of size $(2\ell_{M}+1 )\times 2k_{M}$ centered around the pilot location.  
\subsection{Estimation procedure}
As in other \ac{OTFS} studies e.g.~\cite{health},~\cite{S1D}, channel estimation can be cast as a sparse signal recovery problem
\begin{align}
\label{eq:CS}
\bm{y} &= \bm{\Omega}_{p}\bm{h} + \bm{y}_d + \bm{w}
\end{align}
where $\bm{y}, \bm{w} \in \mathbb{C}^{NM \times 1}$ are defined in \eqref{eq:yvec} and \eqref{eq:ddIO}, 
while the $P$-sparse vector $\bm{h} \in \mathbb{C}^{(l_{M}+1)2k_{M} \times 1}$ that models the channel is given as:
\begin{align}
\label{eq:h}
 \left\{
\begin{aligned}
\bm{h}&\left[\ell_i2k_{M} + k_i\right] = \,h_i, &\text{if }k_i\ge 0 \\
\bm{h}&\left[ (\ell_i+1) 2k_{M} + k_i\right]= \,h_i, &\text{if }k_i< 0 
\end{aligned}
\right. \; \cdot
\end{align}
The columns of the matrix $\bm{\Omega}_{p}\in \mathbb{C}^{NM \times (\ell_{M}+1)2k_{M}}$ are
\begin{equation}
    \label{eq:sigma}
    \bm{\omega}_{i} = (\textbf{F}_N \otimes \textbf{I}_{M})(\ee{}^{-j\frac{2\pi}{NM}k\ell}\bm{\Delta}^{k}\bm{\Pi}^{\ell})(\textbf{F}_N^H \otimes \textbf{I}_{M})\bm{x}_p
\end{equation}
for $i \in \{1, 2,..\, , (\ell_{M}+1)2k_{M}\}$ and ${\bm{x}_p = \text{vec}(\textbf{X}_p)}$, where $\textbf{X}_p$ is the \ac{DD} matrix containing the pilots according to a given design strategy. Each $\bm{\omega}_{i}$ is generated using $\ell$ and $k$ respectively equal to $\ell_i$ and $k_i$ as in the $i$-th element of $\bm{h}$. The terms $\bm{\Omega}_{p}\bm{h}$ and $\bm{y}_{d}$ represent the contributions on $\bm{y}$ of the pilot and the data after being transmitted over the channel. 
Borrowing the nomenclature of \ac{CS} \cite{CS}, $\bm{\Omega}_{p}$ is called \emph{sensing matrix}, and can be leveraged at the receiver to estimate the vector $\bm{h}$. Different from standard \ac{CS}, $\bm{\Omega}_{p}$ cannot be chosen freely as it has to replicate the channel effect and thus the only element we can design is $\bm{x}_p$.

The parameters $\ell_{M}$ and $k_{M}$ are assumed to be known at the transmitter and at the receiver, and can be considered the upper bounds to the channel delay and Doppler spread. The estimation algorithm can exploit the knowledge of these parameters to focus only on the part of the matrix $\textbf{Y}$ that contains contributions related to the pilot. 

When the \ac{S1D} design presented in Sec.~\ref{subsec:pilotsdesign} is adopted, the term $\bm{x}_p$ in \eqref{eq:sigma} is equal to $\text{vec}(\textbf{X}^{(s)}_p)$, and thus the receiver can discard the first $\ell_{M}$ rows of the matrix $\textbf{Y}$ to perform channel estimation. This operation corresponds to the deletion of some specific rows of $\bm{\Omega}_p$. Similarly, when the embedded pilot construction is used we have that $\bm{x}_p= \text{vec}(\textbf{X}^{(e)}_p)$ in \eqref{eq:sigma}. Also in this case, multiple rows of $\bm{\Omega}_{p}$ can be removed since not all elements of $\textbf{Y}$ are needed to estimate the channel.


For both pilot constructions, the resulting sensing matrices show a good behavior from a \ac{CS} point of view since all their columns are independent, implying that the mutual coherence of each of the two matrices is equal to zero. This effect is trivial for the \ac{EP} construction since each column of $\bm{\Omega}_{p}$ will contain only one non zero entry in a different position according to its index. 
Differently from \cite{S1D}, our \ac{S1D} pilot design results in a sensing matrix with zero coherence thanks to the ideal autocorrelation properties of the designed sequence.
As sparse recovery algorithm we employ \ac{OMP}~\cite{OMP}. Note that, given the dimensions of the matrix $\bm{\Omega}_{p}$, the channel estimation problem could also be solved by standard linear methods like \ac{ZF} or \ac{MMSE} but these approaches tend to be complex and yield poorer performance.
It is important to highlight that the channel sparsity, i.e., the number of multi-path channel components, is assumed to be unknown in our setting. The stopping condition of the algorithm is therefore modified to accommodate a threshold on the value of the residual. \ac{OMP} will produce an estimate of $\bm{h}$ that we denote as $\hat{\bm{h}}$.
\subsection{Equalization and iterative estimation approach} 
\label{subsec:iter}
After the execution of the channel estimation procedure, the receiver runs an equalization and demodulation routine to estimate the transmitted data symbols. When data and pilot symbols are non-orthogonal, the pilot contribution has to be canceled from $\bm{y}$ before proceeding with the equalization stage. This operation exploits the channel estimate $\hat{\bm{h}}$ as
\begin{equation}
\label{eq:pilotrem}
\bm{y}_d'= \bm{y} - \bm{\Omega}_{p}\hat{\bm{h}} \; .
\end{equation}
 The vectors $\bm{y}_d'$ and $\hat{\bm{x}}_d$ represent the equalizer input and output respectively.
 We consider as equalizers both the \ac{LMMSE}, which can be derived starting from the input-output relation shown in \eqref{eq:IODD}, and the \ac{MRC} from~\cite{GeneralI/O}. Both algorithms can be implemented more effectively by leaving $\ell_{M}$ empty rows at the bottom of $\textbf{X}_d$ in the form of \ac{ZP}.%
    \footnote{As the Doppler spread grows, the \ac{EP} construction would leave most of the symbols at the bottom of $\textbf{X}_d$ empty in order to maintain data-pilot orthogonality. In such a case, the cost of introducing the \ac{ZP} to the full row leading to a better equalization becomes small. To compare the performance of the different channel estimation methods in a fair manner, we then introduce \ac{ZP} also in our proposed pilot construction \ac{S1D}.} 

Depending on the relative power of data and pilots, the performance of any superimposed channel estimation method may be limited by interference due to the data symbols. To mitigate this phenomenon, we could iterate between channel estimation and equalization. Once a first channel estimate is available, the receiver can estimate the data and perform interference cancellation by removing the estimated data combined with the last channel estimate from the original received signal
\begin{equation}
\label{eq:datarem}
\bm{y}_p'= \bm{y} - \hat{\bm{\Omega}}_{d}\hat{\bm{h}}
\end{equation}
where $\hat{\bm{\Omega}}_{d}$ can be generated through \eqref{eq:sigma} by substituting $\bm{x}_p$ with $\hat{\bm{x}}_d$. At this point, the vector $\bm{y}_p'$ is expected to contain a lower interference from data and it is used as input for the channel estimation procedure that will lead to an enhanced estimate of $\hat{\bm{h}}$. This procedure can be iterated multiple times.




\section{Numerical Results}

In this section, we compare the two pilot constructions and equalization methods presented in Sec.~\ref{sec:pilots}. We consider a multi-path channel with $\ell_{M}=8$, $k_{M}=4$ and $P = 5$. Each propagation component has delay and Doppler shifts uniformly generated from the sets $\{0, .. \,,\ell_M\}$ and $\{-k_M + 1, .. \,, k_M \}$ while the complex gains are $h_i \sim \mathcal{CN}(0,\frac{1}{P})$.
The input data symbols are generated from a 4-\ac{QAM} constellation and are placed in the \ac{OTFS} matrix $\textbf{X}_d$ according to Sec.~\ref{subsec:pilotsdesign}  (see also Fig.~\ref{fig:pilots}). The parameters of the transmitter are set to $M=32$ and $N=16$. In the following, we define the energy devoted to the pilot signal as \emph{pilot energy}, with the understanding that such energy is concentrated on a single DD symbol in the EP case, and spread over multiple in our proposed S1D case. The total energy budget used for the transmission of each OTFS frame is fixed, so that, when $MN$ equal power symbols are generated with all the available energy, their \ac{SNR} equals $15$ dB. In this setting, we vary the energy split between data and pilot while keeping their sum equal to the total available energy budget. The fraction of energy allocated for the pilot varies between $0.05$ and $0.9$. The threshold values used as stopping condition for \ac{OMP} are set as $3\sigma$ for \ac{EP} and $3\sigma + P_d$ for \ac{S1D}, where $P_d$ is the average power of the data symbols.
%
%
%
%
\begin{figure}
    \centering
    \includegraphics[width=0.88\linewidth]{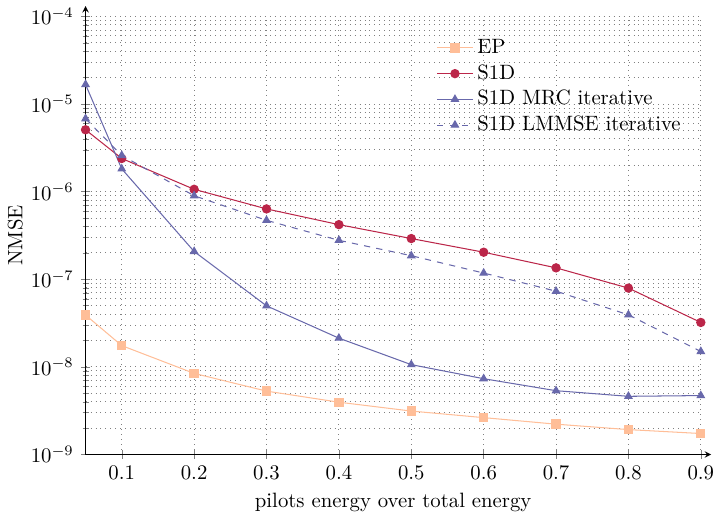}
    \caption{NMSE of channel estimation as a function of the pilot energy.}
    \label{fig:NMSE4QAM}
    \vspace{-0.2cm}
\end{figure}
Fig.~\ref{fig:NMSE4QAM} shows the results for the \ac{NMSE}, defined as $\frac{\parallel\hat{\bm{H}} - \bm{H} \parallel^2_2}{\parallel \bm{H} \parallel^2_2}$ where $\hat{\bm{H}}$ is the estimated \ac{DD} channel matrix in the same form of \eqref{eq:H} but computed with the estimated channel parameters from $\hat{\bm{h}}$. This metric measures the quality of the channel estimate. We note that increasing the pilot symbol energy has a beneficial effect regardless of the pilot construction. For \ac{S1D}, this effect is amplified as the decrease in energy allocated to data also decreases the interference caused on the pilot sequence. The \ac{EP} construction achieves the best \ac{NMSE} across the whole pilots energy range, thanks to data-pilot orthogonality. In order to improve the performance of the \ac{S1D} method, the iterative detection and demodulation technique described in Sec.~\ref{subsec:iter} is considered, fixing the total number of iterations to~$3$.
The \ac{MRC} equalizer shows a remarkable improvement when paired with the iterative procedure, and becomes competitive with the \ac{EP} construction for pilot energy exceeding $0.6$. 
We also note that, for very small pilot energy, the use of the iterative routine may perform worse than the one-shot estimation alone. The reason is that the initial channel estimate may be too imprecise, and causes the incorrect cancellation of data,  hence introducing additional interference on the pilot sequence. 
Finally, for \ac{LMMSE} the addition of the iterative procedure results in a very limited performance gain. The main reason lies in the difficult task of quantifying the amount of residual interference at each iteration, and thus, it appears difficult to find the right regularization parameter of the equalizer.

\begin{figure}
    \centering
    \includegraphics[width=0.88\linewidth]{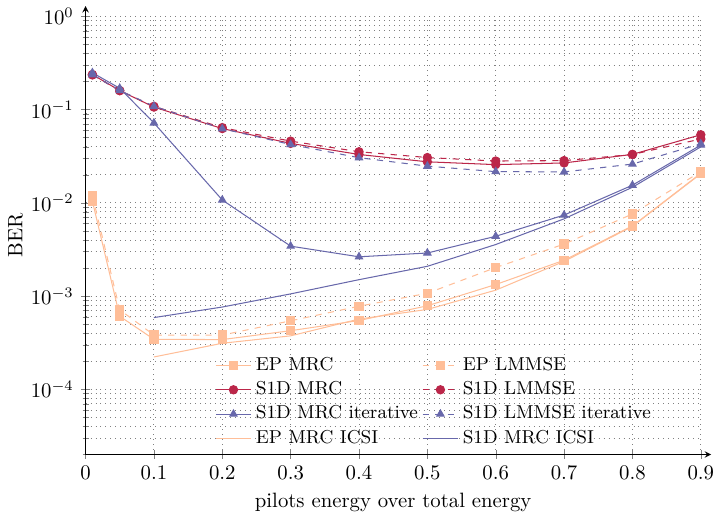}
    \caption{BER as a function of the pilot energy.}
    \label{fig:BER4QAM}
    \vspace{-0.4cm}
\end{figure}
\begin{figure}
    \centering
    \includegraphics[width=0.9\linewidth]{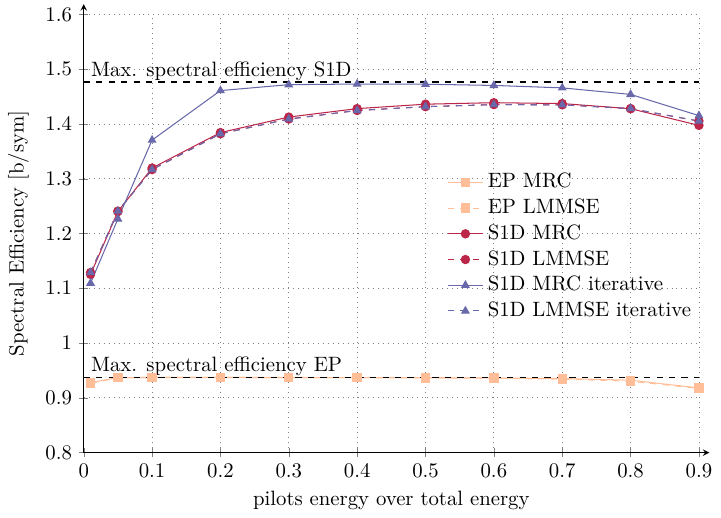}
    \caption{Average spectral efficiency as a function of the pilot energy.}
    \label{fig:TP4QAM}
\end{figure}

Another relevant metric for assessing the pilot construction is the \ac{BER}. Interestingly, different energy splits have to be chosen to minimize the \ac{BER} depending on the selected equalizer and pilot design. In general, as shown in Fig.\ref{fig:BER4QAM}, one needs to find the right balance between a good channel estimate (right-end side of the plot) and a sufficiently high SNR for the data symbols (left-end side of the plot). Allocating exceedingly low energy to the pilot degrades the channel estimate, resulting in poor equalization. On the other hand, when too little power is allocated to the data symbols, the system becomes noise-limited. The minimum \ac{BER} of $\approx 3\cdot 10^{-4}$ is achieved with the \ac{EP} construction, the \ac{MRC} equalizer and $20$\% of energy dedicated to the pilot. This result is tightly followed when employing the \ac{LMMSE} equalizer. The best results achieved with the iterative \ac{S1D} requires a higher pilot energy (i.e., $40\%$ of the total energy) and the best achieved \ac{BER} is $\approx 2.5 \cdot 10^{-3}$ when using \ac{MRC}. The other variants of the \ac{S1D} construction yield worse performance in terms of \ac{BER}, as the interference of data negatively impacts the demodulation. In Fig. \ref{fig:BER4QAM}, we also provide the results with \ac{ICSI} for the different combinations of pilot constructions and equalizers. When sufficient energy is given to the pilots, both the \ac{EP} and \ac{S1D} constructions with iterative \ac{MRC} are very close to the ideal performance. 
The fact that better \ac{BER} results are achieved by the \ac{EP} method reflects the outcome of the \ac{NMSE} analysis. Indeed, the data-pilot orthogonality enables an interference-free estimation of the channel response that can be leveraged to obtain a better performance of the equalizer. This orthogonality, however, comes at the cost of preventing the full exploitation of the available degrees of freedom in the \ac{DD} domain, which must remain empty to avoid interference with the pilots and cannot be used for data transmission. When considering the \ac{S1D} approach, instead, more data symbols can be inserted within each \ac{OTFS} frame. By doing so, the achievable \ac{SE} increases drastically. 
The \ac{SE} is defined as the average number of successfully decoded information bits per transmitted symbol. Note that, in order to avoid inter-frame interference, a \ac{CP} that does not add information content is pre-pended to transmitted signal thus increasing the overall length of the frame. This CP penalty is only considered for the \ac{S1D} construction since for the \ac{EP} one the presence of the full \ac{ZP} guard in \ac{DD} domain already prevents inter-frame interference.
The \ac{SE} results are presented in Fig.~\ref{fig:TP4QAM}, where \ac{S1D} shows a $\approx 57$\% gain when compared with the \ac{EP} method. 
\begin{figure}
    \centering
    \includegraphics[width=0.85\linewidth]{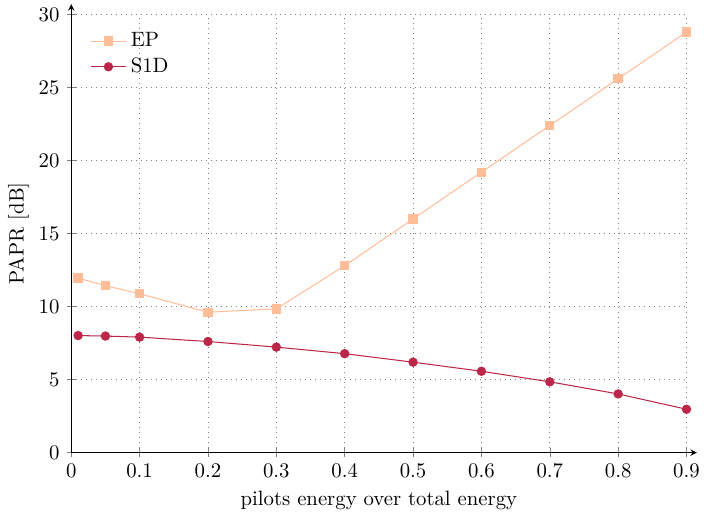}
    \caption{PAPR as a function of the pilots energy.}
    \label{fig:PAPR4QAM}
    \vspace{-0.4cm}
\end{figure}
In Fig.~\ref{fig:PAPR4QAM} we show the results in terms of \ac{PAPR}. The \ac{S1D} construction provides the best performance regardless of the pilot energy allocation. With this type of construction, the \ac{PAPR} is decreasing in function of the pilots energy as the main contribution is given by the data signal. For the \ac{EP} placement, instead, when most of the energy is allocated to the pilot the \ac{PAPR} grows very large. It is worth noting that the \ac{EP} construction achieves the best \ac{BER} and \ac{PAPR} for the same energy split of $0.2$. We leave for future study whether this holds also for other channel models and \ac{OTFS} frame sizes. 
Considering the best operating points of both pilot constructions in terms of \ac{BER}, the use of the \ac{S1D} method enables a \ac{PAPR} reduction of $\approx\!3$~dB. 

An apparent result that appears from this case study is that the performance of the communication system has a strong dependency on the selected energy allocation. The intrinsic relationship between performance metrics and the instantaneous parameters (including channel characteristics, noise level, and system numerology) is expected to be non-trivial. In this regard, further research is needed to define a strategy to split the available energy resources a-priori.

From the point of view of complexity, the choice among \ac{EP} and \ac{S1D} introduces a possible difference in the cost of executing the \ac{OMP} algorithm at the receiver side. In particular, the number of rows of the sensing matrix $\bm{\Omega}_p$ for \ac{EP} and \ac{S1D} are given as $(\ell_M+1)2k_M$ and $(M-\ell_M)2k_M$ respectively.  When $M \approx \ell_M$ the choice on the pilot constructions does not affect the complexity cost. Note that,  since the columns of $\bm{\Omega}_p$  only represent integer \ac{DD} shifts of the transmitted pilot,  for the considered constructions the resulting sensing matrices show ideal coherence properties, meaning that all columns of $\bm{\Omega}_p$ are linearly independent. In this specific setting, the algorithm OMP can be equally implemented without the need of the least square operation thus avoiding the need of any matrix inversion. Finally, a linear increase in complexity appears when iterative channel estimation and demodulation are employed. In this case, complexity cost is proportional to the number of iterations of interference cancellation.

\begin{figure}[t]
    \centering
    \includegraphics[width=0.86\linewidth]{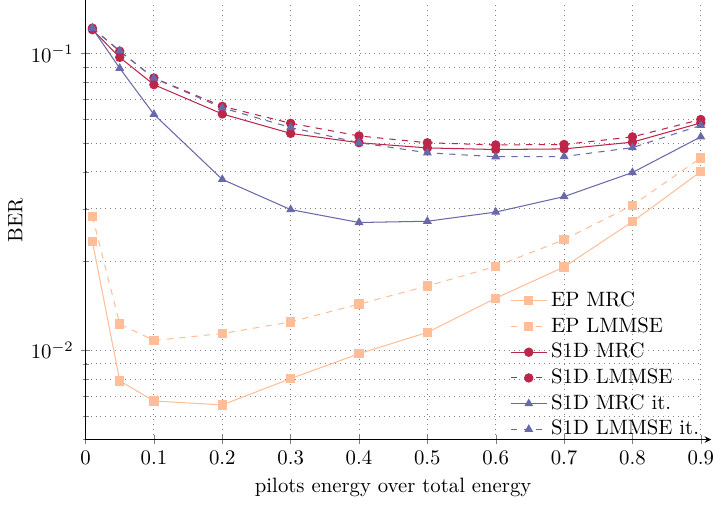}
    \caption{BER as a function of the pilot energy, with input symbols from the 16-\ac{QAM} constellation.}
    \label{fig:BER16QAM}
\end{figure}
\begin{figure}[t]
    \centering
    \includegraphics[width=0.86\linewidth]{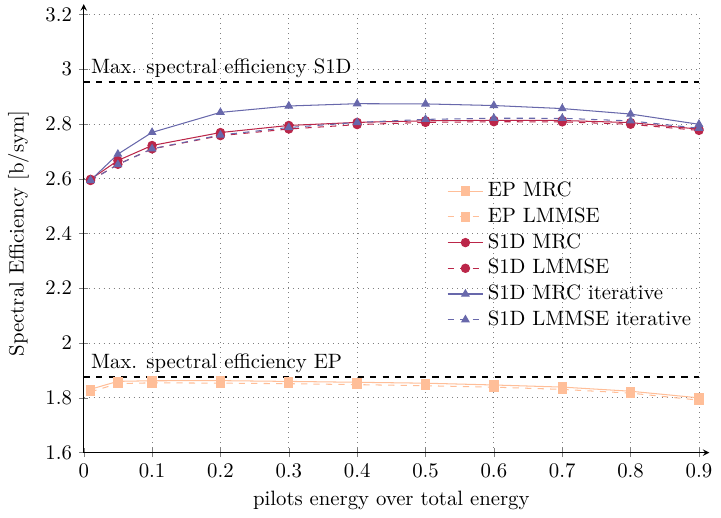}
    \caption{\ac{SE} as a function of the pilot energy, with input symbols from 16-\ac{QAM} constellation.}
    \label{fig:TP16QAM}
    \vspace{-6pt}
\end{figure}
As a final experiment, we consider the use of 16-\ac{QAM} to generate the input data symbols. By increasing the modulation order while maintaining the average symbol power fixed, one trades \ac{BER} for increased \ac{SE}, as shown in Fig.~\ref{fig:BER16QAM} and Fig.~\ref{fig:TP16QAM}. Even when no \ac{PAPR} requirements are considered, the \ac{S1D} construction still enables the successful reception of a higher number of bits per \ac{OTFS} frame. When instead an upper bound on the \ac{PAPR} is considered,  only a small energy split range can be considered to fulfill the requirement. When the \ac{PAPR} limit is tight, an energy back-off might be necessary and in this latter case, we expect the \ac{S1D} to better scale given its superiority on this performance criterion.

When the objective is to maximize the \ac{SE}, the target \ac{BER} should also be considered. If this requirement allows the use of higher-order modulations for both pilot constructions, then the proposed \ac{S1D} method delivers the best performance. However, if the target error probability limits larger constellation sizes only for the superimposed data-pilot construction, the orthogonal \ac{EP} method should be preferred. Overall, depending on the specific \ac{BER} and \ac{PAPR} targets, different data-pilot designs should be chosen to maximize the \ac{SE}.


\section{Conclusions}
In this paper, we proposed a novel superimposed pilot construction for \ac{OTFS} modulation, designed to improve the \ac{SE} and lower the \ac{PAPR}. This construction shows favorable properties when channel estimation is cast as a \ac{CS} problem. The role of the data-pilot orthogonality in the \ac{DD} domain is discussed and evaluated via numerical simulations, where we compare the performances of two representative systems. 
We show that, when iterative estimation and successive cancellation are employed along with our proposed method, a small performance penalty in terms of \ac{NMSE} of the channel estimate and hence \ac{BER} can be exchanged for significant gains in terms of \ac{PAPR} and \ac{SE} compared to data-orthogonal designs.

\bibliographystyle{IEEEtran}
\bibliography{main}

\begin{acronym}
        \acro{2SRA}{two-step random access}
        \acro{4-QAM}{4-quadrature amplitude modulation}
        \acro{16-QAM}{16-quadrature amplitude modulation}

        \acro{4SRA}{four-step random access}
        \acro{5GNR}{5G New Radio}
        \acro{ADC}{analog-to-digital converter}
        \acro{AWGN}{additive white Gaussian noise}
        \acro{BER}{bit error rate}
        \acro{BLER}{block error rate}
        \acro{BP}{belief propagation}
        \acro{BS}{base station}
        \acro{CCS}{coded compressed sensing}
        \acro{CAZAC}{constant amplitude zero autocorrelation}
        \acro{c.c.u.}{complex channel use}
        \acro{CP}{cyclic prefix}
        \acro{CRDSA}{contention resolution diversity slotted Aloha}
        \acro{CS}{compressed sensing}
        \acro{CSA}{coded slotted Aloha}
        \acro{c.u.}{channel use}
        \acro{DAC}{digital-to-analog converter}
        \acro{DD}{delay-Doppler}
        \acro{DFT}{discrete Fourier transform}
        \acro{DZT}{discrete Zak transform}
        \acro{EP}{embedded pilot}
        \acro{ICI}{inter-carrier interference}
        \acro{ICSI}{ideal channel state information}
        \acro{IDMA}{interleaver division multiple access}
        \acro{IDFT}{inverse discrete Fourier transform}
        \acro{IDZT}{inverse discrete Zak transform}
        \acro{IFFT}{inverse fast Fourier transform}
        \acro{IoT}{Internet of Things}
        \acro{IRSA}{irregular repetition slotted Aloha}
        \acro{ISI}{inter-symbol interference}
	    \acro{LDPC}{low-density parity-check}
        \acro{LLR}{log-likelihood ratio}
        \acro{LMMSE}{linear minimum mean squared error}
        \acro{LTE}{Long Term Evolution}
        \acro{MAC}{multiple access}
        \acro{MIMO}{multiple-input multiple-output}
        \acro{MMSE}{minimum mean square error}
        \acro{MPR}{multi-packet reception}
        \acro{MRC}{maximal-ratio combining}
        \acro{MTC}{machine-type communication}
        \acro{MTO}{many-to-one}
        \acro{NB-IoT}{Narrowband IoT}
        \acro{NMSE}{normalized mean squared error}
        \acro{NR}{new radio}
        \acro{OFDM}{orthogonal frequency-division modulation}
        \acro{OMP}{orthogonal matching pursuit}
        \acro{OTFS}{orthogonal time frequency space}
        \acro{OTO}{one-to-one}
        \acro{PAM}{pulse-amplitude modulation}
        \acro{PAPR}{peak-to-average power ratio}
        \acro{PBCH}{physical broadcast channel}
        \acro{PDSCH}{physical downlink shared channel}
        \acro{PO}{PUSCH occasion}
        \acro{PRACH}{physical random access channel}
        \acro{PRB}{physical resource block}
        \acro{PSS}{primary synchronization signal}
        \acro{PUPE}{per-user probability of error}
        \acro{PUSCH}{physical uplink shared channel}
        \acro{QPSK}{quadrature phase shift keying}
        \acro{QAM}{quadrature amplitude modulation}

        \acro{RA}{random access}
        \acro{RCP}{reduced cyclic prefix}
        \acro{RCU}{random coding union}
        \acro{SB-IDMA}{sparse block interleaver division multiple access}
        \acro{SCL}{successive cancellation list}
        \acro{SCS}{sub-carrier spacing}
        \acro{SE}{spectral efficiency}
        \acro{SIC}{successive interference cancellation}
        \acro{SNR}{signal-to-noise ratio}
        \acro{SPARC}{sparse regression code}
        \acro{S1D}{superimposed pilot}
        \acro{SSS}{secondary synchronization signal}
        \acro{TA}{time advance}
        \acro{TF}{time-frequency}
        \acro{TBS}{transport block size}
        \acro{TIN}{treat-interference-as-noise}
        \acro{UT}{user terminal}
        \acro{UMAC}{unsourced multiple access}
        \acro{ZC}{Zadoff-Chu}
        \acro{ZF}{zero forcing}
        \acro{ZP}{zero padding}
\end{acronym}

\end{document}